\documentstyle[12pt,aps,epsf,preprint,tighten]{revtex}

\begin{document}
\draft
\title{Array-induced collective transport in the Brownian motion of
coupled nonlinear oscillator systems}

\author{Zhigang Zheng$^{1,3}$ Bambi Hu$^{1,4}$ and Gang Hu$^{2,3}$}
\address{
$^{1}$ Department of Physics and Center for Nonlinear Studies, Hong Kong 
Baptist University, Kowloon Tong, Hong Kong, China  \\
$^{2}$ Center of Theoretical Physics, Chinese Center of Advanced Science
and Technology (World Laboratory), Beijing 8730, China \\
$^{3}$ Department of Physics, Beijing Normal University, Beijing 100875,
 China \\
$^{4}$ Department of Physics, University of Houston, Houston TX 77204, USA}
\maketitle

\begin{abstract}
Brownian motion of an array of harmonically coupled particles subject to
a periodic substrate potential and driven by an external bias is
investigated. In the linear response limit (small bias), the coupling
between particles may enhance the diffusion process, depending on the
competition between the harmonic chain and the substrate potential. An 
analytical formula of the diffusion rate for the single-particle case
is also obtained. In the
nonlinear response regime, the moving kink may become phase-locked to its
radiated phonon waves, hence the mobility of the chain may decrease as one
increases the external force.
\end{abstract}


\section{Introduction}

Brownian motion of particles subject to periodic substrate potentials and
external forces gained great interest due to its wide applications and
practical importance in connection with transport processes in many fields
including damped pendula, superionic conductor, Josephson tunneling
junction, vortex motion in high-T$_c$ oxide superconductors, phase-locked
loop, rotation of dipoles, charge-density wave, dislocation, and so on
[1-7]. Although the collective transport of coupled nonlinear oscillators has
been studied recently for the two-dimensional case in relation to studies on 
adsorbate islands and monolayer films on surfaces [8,15,17], explorations of collective 
behaviors for the one-dimensional coupled nonlinear oscillators are still
fundamentally important in understanding many physical systems. 
In dimensionless form, the Langevin equation in describing the 1D case 
might be written as

\begin{equation}
\label{1}\stackrel{..}{x}_j+\gamma \stackrel{.}{x}_j+d\sin \left( \frac{2\pi
x_j}b\right) =\sum\limits_{i=1}^{N}\frac{\partial V(x_j,x_i)}{\partial x_j}%
+F_j(t)+\xi _j(t), 
\end{equation}
where $x_j$ represents the coordinate of the {\it i}-th particle and $%
\stackrel{.}{x}_j=dx_j/dt$ its corresponding velocity, $\gamma $ is the
friction coefficient, $d$ is the height of the periodic potential, 
and $b$ denotes the period of the substrate potential. 
$V(x_j,x_i)$ reflects the interaction between 
the {\it j}-th and the {\it i}-th particle. $F_j(t)$ and $\xi _j(t)$ denote the
external driving force and the thermal fluctuation induced random force on
the {\it j}-th particle, respectively. In this paper we focus on the case when  the 
external 
driving is uniformly constant, i.e., $F_j(t)=F$. The thermal noise is frequently
assumed to be a both spatially and temporally uncorrelated Gaussian-type
one: 
\begin{equation}
\label{2}\left\langle \xi _j(t)\right\rangle =0,\quad \left\langle \xi
_i(t)\xi _j(t^{\prime })\right\rangle =2\gamma k_BT\delta _{i,j}\delta
(t-t^{\prime }), 
\end{equation}
where $k_B$ is the Boltzmann constant, and $T$ is the environmental
temperature. When the interaction among particles is very weak, in many cases
one may approximately treat the above problem in terms of Brownian motion of a
single particle under a biased periodic potential. The corresponding
Fokker-Planck equation then reads [1] 
\begin{equation}
\label{3}\frac{\partial W}{\partial t}=-\stackrel{.}{x}\frac{\partial W}{%
\partial x}+\gamma \frac \partial {\partial \stackrel{.}{x}}\left( \stackrel{%
.}{x}W+k_BT\frac{\partial W}{\partial \stackrel{.}{x}}\right) . 
\end{equation}
Here $W(x,\stackrel{.}{x},t)$ denotes the probability distribution of the
particle, where the indices of particles are omitted. This equation is
sufficiently complicated to be analytically solved so that a closed analytical
solution in describing the diffusion process is far from obtainable. Based on
the inverse-friction expansion method, we recently gave an analytical
perturbative solution of the mobility for the single-particle case valid for arbitrary
friction cases [9]. When the interaction among particles can no longer be
ignored, one has to treat the coupled case (1). Interactions between particles 
introduce new time scales, which
leads to more complicated phenomena, and furthermore an analytical treatment
of the corresponding Fokker-Planck equation is almost impossible [3]. One of
the simplest models in describing the competition between the coupling and
the substrate potential is the well-known Frenkel-Kontorova system [10],
which describes a chain of particles with nearest-neighboring harmonic
couplings and subject to a periodic potential, where the interaction is
reduced to 
\begin{equation}
\label{4}V(x_j,x_{j-1})=\frac 12K(x_j-x_{j-1}-a)^2, 
\end{equation}
where $K$ and $a$ are the coupling strength and the static length of the
spring, respectively. Now there are two competing lengths: $a$ and $b$. The
winding number (or the frustration) is defined as $\delta =b/a$, which may
strongly affect the spatial configuration of the system. During the past few
years, the FK model was applied to investigations of the ground state of
competing systems, and the commensurate-incommensurate (CI) phase
transitions were found and theoretically explored [11]. The theory developed
by Aubry stands as one of the deepest achievements in theoretical
comprehension of the physics of modulated phases [12]. Dynamics of the FK
chain was also explored in relating to many fields, such as charge-density
wave (CDW), nano-tribology and surface problem, self-organized criticality
(SOC) and Josephson-junction arrays (JJA) and ladders (JJL) [13, 14]. In the
following investigations, we simply set $b$ to be $2\pi $. Then the Langevin
equation connected to the FK chain can be followed from (1):

\begin{equation}
\label{5}\stackrel{..}{x}_j+\gamma \stackrel{.}{x}_j+d\sin
x_j=K(x_{j+1}-2x_j+x_{j-1})+F+\xi _j(t). 
\end{equation}
One may notice that the
static length of the spring $a$ does not enter (5), but it may play a very
significant role in describing the motion of the coupled systems. The
corresponding Fokker-Planck equation reads 
\begin{equation}
\label{6}
\frac{\partial W}{\partial t}=\sum\limits_{j}\left[ -\stackrel{.}{x}_j\frac{%
\partial W}{\partial x_j}+\frac{\partial U}{\partial x_j}\frac{\partial W}{%
\partial x_j}+\gamma \frac \partial {\partial \stackrel{.}{x}_j}\left( 
\stackrel{.}{x}_jW+k_BT\frac{\partial W}{\partial \stackrel{.}{x}_j}\right)%
\right] , 
\end{equation}
where $W=W(\left\{ x_j\right\} ,\left\{ \stackrel{.}{x}_j\right\} ,t)$ is
the joint probability distribution function, and here we use the total
potential $U(\left\{ x_j\right\} )$, and for the case of the FK model 
$ U(\left\{ x_j\right\} )=\sum\limits_{j}[d(1-\cos x_j)+\frac
12K(x_{j+1}-x_j-a)^2]$. In the high-friction limit one may obtain the 
Smoluchowski equation by averaging velocities, which can be written as 
\begin{equation}
\label{7}\frac{\partial P}{\partial t}=\frac 1\gamma \sum\limits_{j}%
\frac \partial {\partial x_j}\left( \frac{\partial U}{\partial x_j}P+k_BT 
\frac{\partial P}{\partial x_j}\right) , 
\end{equation}
where $P(\left\{ x_j\right\} ,t)=\int W(\left\{ x_j\right\} ,\left\{ 
\stackrel{.}{x}_j\right\} ,t)\prod\limits_{j}d\stackrel{.}{x}_j$ is the
reduced probability distribution for only spatial variables. In many cases
one is interested in the mobility of the chain: 
\begin{equation}
\label{8}\mu =\frac{<v>}F, 
\end{equation}
where $<.>$ includes both time and particle averages. This definition was
introduced in discussions of Brownian motion of a single particle in the
biased periodic potential. We will give a unified solution for the single
particle case in the following discussion. Analytical investigations of the
coupled case may be more difficult, therefore we will discuss this problem
mainly in terms of numerical simulations. The {\it collective diffusion
coefficient}, which is relevant for studying commensurability effects [15],
is given by the linear part of the mean-square displacement: 
\begin{equation}
\label{9}D=\lim _{t\rightarrow \infty }\frac 1{2Nt}\sum\limits_{i,j}\left\langle
[x_i(t)-x_j(0)]^2\right\rangle . 
\end{equation}
This coefficient is connected to the mobility in the small force limit ({\it %
linear response}) [1] by 
\begin{equation}
\label{10}D(T,K,\delta )=k_BT\lim _{F\rightarrow 0}\mu (F,T,K,\delta ). 
\end{equation}
One may notice that the diffusion coefficient relates to the temperature,
the coupling strength and the frustration. In particular, the dependence on
the coupling strength and the frustration does not happen for the single
particle case. Studies of their dependences are very interesting and
significant. In the following discussions, we will investigate both the
linear and the nonlinear response regimes. We find that in the limit of
linear response, the coupling between particles can greatly increase the
diffusion rate, which may be important in realistic experiments and
technologies. Additionally, we shall investigate the commensurability effects. In
the vicinity of the golden mean $\delta =\delta _G=(\sqrt{5}-1)/2$, 
the chain possesses the maximum diffusion
coefficient. In the nonlinear response regime, the transport is dominated by
the strong resonant behavior due to the competitive phase-locking between the
travelling wave and its radiated linear phonons. This resonance may lead to
the suppression of the mobility. In the following investigations, we mainly
perform numerical simulations for the coupled case. The fourth-order
Runge-Kutta integration algorithm is applied and the time step is adjusted
according to the numerical accuracy. Periodic boundary conditions are
applied, i.e., $x_{j+N}(t)=x_j(t)+2\pi M$, where $M$ is an integer that
counts the net number of kinks trapped in the ring, therefore the
frustration is $\delta =M/N$ and the spring constant will be $a=2\pi \delta $.
Throughout the paper $ \gamma =0.1$ and $d$ is set to be 1.

\section{ Linear response: enhancement of the diffusion}

In Fig.1(a), we give the relation between the diffusion coefficient $D$ and
the coupling strength $K$ for the incommensurate case (e.g., the gold mean case
$\delta =\delta _G$)
and for different temperatures. The first phenomenon we observe is that the
relation between the diffusion coefficient and the coupling strength is not
monotonic. For weak coupling strengths, the diffusion process is suppressed;
When $K>K_{c1}$ ({\it the first threshold}), the diffusion coefficient
begins to increase as one increases the coupling strength. Near this
critical value, the diffusion coefficient $D$ is found to 
obey the following relation: 
\begin{equation}
\label{11}D=D_0(T)\left| K-K_{c1}\right| ^{\alpha (T)}, 
\end{equation}
where the scaling exponent $\alpha (T)$ decreases with increasing the
temperature $T$. At a {\it second threshold} $K_{c2}$, $D$ begins exceeding
the single-particle value ($K=0$, the non-interacting case), exhibiting an {\it 
array-enhanced diffusion process}. For the moderate coupling strength, we find that 
the $D-K$ relation obeys the power law: 
\begin{equation}
\label{12}D\propto K^{\beta (T)}, 
\end{equation}
where the scaling exponent scales with the temperature as $\beta (T)\sim
T^{-1/2}$, i.e., it decreases with increasing the temperature. At the high
coupling constant, the diffusions coefficient may saturate to a value much higher
than the single-particle case. This behavior is
very interesting, because one must overcome a {\it threshold coupling}
before one can get a higher diffusion rate. Additionally, $K_{c1}$ and 
$K_{c2}$ increases with the temperature, i.e., for higher temperatures, one
must overcome stronger coupling thresholds to get higher diffusion rates.
The result indicates that if one introduces some coupling between the
particles (elements) then a higher diffusion rate than uncoupled systems can
be achieved. In many realistic applications one hopes that the diffusion
process can be improved as fast as possible. Our exploration indicates that
for the incommensurate case the coupling between particles may enhance the
diffusion process.

The above behavior can be heuristically interpreted, which is a typical
consequence of{\it \ the competition between order and disorder}. The
mechanism of order comes from the coupling among particles, where the
coupling tends to organize the chain to move in a
collective way; the mechanism of disorder is the thermal fluctuation, which
tends to destruct the ordered motion. For very weak couplings, the thermal
noise may dominate, thus particles cannot organize themselves very well to
diffuse collectively. In this case interactions between elements introduce
another resource of the dissipation that leads to higher friction [18]. When the
coupling between particles increases and exceeds a threshold, then the role
of disorder may be supressed, and particles can gradually move collectively,
leading to high diffusion rates. At higher temperatures, the noise produces
much more disorder, hence the chain needs a stronger coupling to organize
the collective diffusion.

All the above discussions are valid for the incommensurate case. In
Fig.1(b), we give the numerical result of the diffusion coefficient against
the winding number (frustration) $\delta $ for different temperatures, and
the coupling strength $K=1.0$, which strongly enough for the chain to
diffuse. It can be found that for high temperatures, the commensurability
effect is not very significant. Because the curve is symmetric about $1/2$,
we only discuss the range $\delta \in [0,1/2]$. For small winding numbers,
the relation is a linear one (see Fig.1(b)): 
\begin{equation}
\label{13}D=D_0T\delta , 
\end{equation}
where the slope is proportional to the temperature. This indicates that for
small frustrations, the diffusion is suppressed by introducing the coupling
among particles. The curve becomes flat when $\delta $ 
further increases. This result indicates that although the particles are coupled, for small
frustrations, the diffusion process is still very slow. In this case
couplings between particles act as an additional source of dissipation [18].
This suppression of the diffusion is because the PN barrier is high (for the coupled 
case the motion of the chain is dominated by the moving kink, as pointed out below. 
The Peierls-Nabarro barrier corresponds to the barrier for the kink
translation along the chain. It is also the minimum energy necessary to move
the kink along the chain.), and the
collision of particles with the substrate potential plays a more significant
role, thus one needs a higher activated energy to overcome the PN barrier.
Commensurate effect is clearly shown when the
temperature decreases. For the case $k_BT=0.25$, we observe that the
diffusion rate reaches a maximum value at approximately $\delta =1/3$ and 
$2/3$. These two values are very close to the golden mean value $\delta _G=( 
\sqrt{5}-1)/2$ and $1-\delta _G$. In order to get a higher diffusion, one
should choose winding numbers near the golden mean. This result can be
applied to the dynamics of Josephson junction arrays, where the
frustration can be altered by changing the magnetic field strength [19].

The relation between $D$ and $T$ is not trivial. We first discuss the
non-interacting case. For the very small damping constant and high enough
temperature case, the diffusion coefficient was approximately obtained as [1] 
\begin{equation}
\label{14}D=\frac{\pi k_BT}{2\gamma }\exp (-\frac{2d}{k_BT}), 
\end{equation}
i.e., the diffusion rate increases with the temperature. In the high damping
and low temperature limit, the diffusion is dominated by thermally activated
hoppings, then one has the following formula [1]: 
\begin{equation}
\label{15}D=\frac{k_BT}\gamma \left[ I_0(\frac d{k_BT})\right]
^{-2}\rightarrow \frac{2\pi d}\gamma \exp (-\frac{2d}{k_BT}), 
\end{equation}
where $I_n(x)$ is the modified Bessel function. one may find the same
Boltzmann factor (Arrhenius form) $\exp (-2d/k_BT)$ as the low-damping case,
except the prefactor. The common factor comes from the
thermal-fluctuation-induced hopping effect. In general cases, we are still
able to derive a unified formula of the diffusion rate. We investigated the
mobility and got a successively perturbative solution of the mobility [9].
By using the relation between the mobility and the diffusion coefficient
(10), in the linear response limit, $F\rightarrow 0$, we have the formula: 
\begin{equation}
\label{16}D=\frac{(2\pi )^2k_BT}{\gamma [\Omega (d,0)\Delta (d,0)-\Lambda
(\gamma ,T,d,0)]}, 
\end{equation}
where $\Omega (d,F),\Delta (d,F)$ and $\Lambda (\gamma ,T,d,F)$ are given by
$$
\Omega (d,F)=\int_0^{2\pi }\exp [f(x)-Fx]dx, 
$$
$$
\Delta (d,F)=\int_0^{2\pi }\exp [-f(x)+Fx]dx, 
$$
\begin{equation}
\label{17}\Lambda (\gamma ,T,d,F)=\int_0^{2\pi }dx\exp
[-f(x)+Fx]\int_0^xq(\xi )\exp [f(\xi )-F\xi ]d\xi , 
\end{equation}
where $f(x)=-dcosx$. The kernel function $q(x)$ can be obtained from the
following operation (see [9] for a detailed discussion): 
\begin{equation}
\label{18}q(x)=\frac 1c\gamma \stackrel{\wedge }{\bf H}c, 
\end{equation}
where $c$ is a constant and $\stackrel{\wedge }{\bf H}$ is a
continued-fraction operator acting on the constant $c$: 
\begin{equation}
\label{19}\stackrel{\wedge }{\bf H}=\stackrel{\wedge }{I}-\frac 1{\gamma ^2}%
\stackrel{\wedge }{K}^{+}\frac 1{\stackrel{\wedge }{I}-\frac 1{2\gamma ^2}%
\stackrel{\wedge }{K}^{+}\cdot \cdot \cdot \stackrel{\wedge }{K}^{-}}%
\stackrel{\wedge }{K}^{-}, 
\end{equation}
where $\stackrel{\wedge }{I}$ is the unit operator and $\stackrel{\wedge }{K}%
^{+},\stackrel{\wedge }{K}^{-}$ are given as%
$$
\stackrel{\wedge }{K}^{+}=\sqrt{k_BT}\frac \partial {\partial x}, 
$$
\begin{equation}
\label{20}\stackrel{\wedge }{K}^{-}=\stackrel{\wedge }{K}^{+}+\frac 1{\sqrt{%
k_BT}}\left[ \frac{df(x)}{dx}-F\right] . 
\end{equation}
This is a unified formula of the diffusion coefficient valid for arbitrary
damping and temperature cases.

When the particles are coupled in an array way, they will move collectively,
leading to the wave motion. In this case, the diffusion is mainly related to
the kink motion (travelling wave) [14, 16]. The kink describes the minimally
possible, topologically stable compression of the commensurate structure.
The kink is a quasiparticle, characterized by an effective mass, rest energy
and the Peierls-Nabarro (PN) amplitude $E_{PN}$. In the low temperature
case, the diffusion rate can be given as [17] 
\begin{equation}
\label{21}D=D_0\exp (-\frac{E_{PN}}{k_BT}), 
\end{equation}
where $D_0$ is the prefactor that scales with $E_{PN}$ as $D_0\propto
E_{PN}^{1/2}$ for the low-damping case and $D_0\propto E_{PN}$ for the
high-damping case. In the strong coupling case, one may have 
\begin{equation}
\label{22}D\approx Ck_BT\left[ 1-\frac 18\left( \frac{E_{PN}}{k_BT}\right)
^2\right] . 
\end{equation}
Eq. (22) indicates that there are two
competing terms. For very high temperatures, the diffusion rate may be
dominated by $D_0$. In the low-temperature regime, the two terms may compete
and the exponential term plays an important role. Because the PN barrier
closely relates to the commensurability effect, the diffusion rate will be
strongly affected by the commensurability of the system. It may be easily
found that for both the single particle and the coupled particle systems, an
Arrhenius factor always exists, except that {\it for the single case the
factor relates to the potential barrier, but for the chain it relates to the
PN barrier.}

\section{Nonlinear response: Suppression of the mobility}

When one increases the external driving force, the transport behavior
becomes a drift one. In this case one usually uses the mobility $\mu
=\left\langle v\right\rangle /F$ to describe the transport process of the
system. In Fig. 2(a) and (b), we give the the numerical $\mu -F$ relation
for $\delta =3/$$8$ and $1/8$, respectively. The mobility of the single
particle for $k_BT=0.50$ is also plotted to make a clearer comparison. 
The first result one
may clearly observe is that when the driving force is small, the mobility of
the chain is much higher than that of the single-particle case. As one
increases the force, the mobility decreases. These two phenomena are 
much different from the
single-particle case (see the single-particle line in Fig.2). Another novel
behavior different from the single-particle case is that at the left side
of the decreasing lines(e.g., different lines between 0.15 and 0.4 in Fig.2(a) 
corresponding to different temperatures, see the left side $F=0.15$), 
the mobility decreases when one increases the temperature (see, 
for example, $F=0.15$ in the inset of Fig.2(a), the mobilities corresponding 
to higher temperatures are smaller.).
In the middle of the decreasing lines (e.g., $F=0.25$ in Fig.2(a)), 
the mobility is almost unaffected by the increasing
temperature. At the right hand of the decreasing lines, the mobility increases with
increasing the temperature, which is a natural consequence (Similar phenomenon 
could also occur at approximately $F=0.1$). The above
anomalous phenomena are consequences of the interaction among
particles. In fact, one may study the noiseless case to find this intrinsic
reason. In this case it is found [20-22] that the motion of the
chain is dominated by the moving localized kink, i.e., a distorted
travelling wave that is composed of a moving kink and the oscillating linear
wave around it. In the underdamped case, the moving kink may become
phase-locked to its radiated phonons [14]. We have obtained a mean-field
formula of the resonance-velocity spectrum (see [21] for a detailed derivation): 
\begin{equation}
\label{23}v(m_1,m_2)=\frac{m_2}{m_1}\sqrt{\beta +4K\sin {}^2(\frac{m_2\delta
\pi }{m_1})}.
\end{equation}
The resonance is denoted by a pair of integers $(m_1,m_2)$. Here $\beta=<\sum_{j=1}%
^{N}\frac {1}{N} \cos(x^{*}(t))> $ is
called the {\it contraction factor}, which describes the collective effect
of the chain (kink). Here ${x^{*}(t)}$ denotes the steady state of the system (5), and 
$<.>$ represents the time average. In Fig.3(a), we give the relation between 
$\beta $ and $\delta $ for $F=0$. In this case, commensurate effect can be clearly
observed, where some rational peaks correspond to commensurate cases 
$\delta =0,1,1/2,1/3,2/3,...$. In
Fig. 3(b), we plot the contraction factor $\beta $ vs. the external driving 
$F$. The dashed horizontal line is the value for $F=0$. It is clearly
illustrated that $\beta $ varies with $F$ in a non-monotonic way. The value
of $\beta $ for $F\neq 0$ is higher than that of $F=0,$ i.e., the chain is
further contracted when one increases $F$. At $F\approx 0.47$, $\beta $
suddenly drops off to a negative value. The compressed topology of the chain
(kink) becomes extended, indicating a transition to a new state. This
transition is also shown in Fig. 2(a) and (b), where the mobility jumps to a
much higher value. We called this regime with a high mobility the
''high-velocity regime''. This effect will be described in another work,
which is beyond our present interest.

The resonance behavior between the travelling wave and its radiated phonons
remains robust even when finite temperature effect (thermal noise) is
considered. The mode locking behavior leads to resonant steps of the
averaged velocity of the chain when one adiabatically increases the external
force, thus a decrease of the mobility with increasing forces can be
expected. When the driving force is increased along a resonant step, the
input energy is mainly consumed for amplifying the excited linear phonon and
even exciting new linear modes, while the average velocity of the chain
remains nearly constant. Thus the coupling between particles here acts as
an additional source of dissipation. Due to the amplification of the linear
wave, the resonance will be gradually unstable and eventually destructed
when one increases the external force. As one increases the temperature, the
excited linear waves will be further amplified. At the right end of a
resonant step these very strong linear waves would destruct the resonances,
thus the mobility increases. This is the so-called "{\it smearing effect}". 
At the left right hand of the resonance, the noise can only 
slightly affect the resonance, hence the coupling between particles may stabilize the
resonance. In this case the noise tends to drive the resonance state to a
lower resonance state, thus the mobility decreases when the temperature is
increased. This interprets the observation reported above. To get a deeper 
understanding, in Fig.4, we give the relation between the mobility and the
temperature for different external forces $F=0.05$ and $0.10$ at $\delta =3/8$. 
$F=0.10$ corresponds
approximately to the left hand of the resonance and $F=0.05$ the middle of
the resonance. It can be clearly shown that in the middle of the resonance,
the mobility keeps nearly constant; while at the left hand, the mobility
decreases with the temperature.

In conclusion, in this paper we investigated the collective 
transport behavior of the underdamped 
Frenkel-Kontorova chain under a constant external driving force and the influence
of the environmental noise, and found complicated behaviors. Results reported in
this paper should be valuable of applications in many physical cases, such as 
charge-density waves, Josephson junction arrays and ladders and tribology. Some 
results need a theoretical exploration, and studies along this line are 
currently in progress.

\bigskip\ \bigskip\ 

One of the authors(Zheng) thanks all the colleagues in Center for Nonlinear
Studies of Hong Kong Baptist University for many valuable discussions. This
work is supported in part by the Research Grant Council RGC and the Hong
Kong Baptist University Faculty Research Grant FRG. Support by
the National Natural Science Foundation of China is acknowledged.

\begin{figure}
\epsfxsize=8cm
\epsfbox{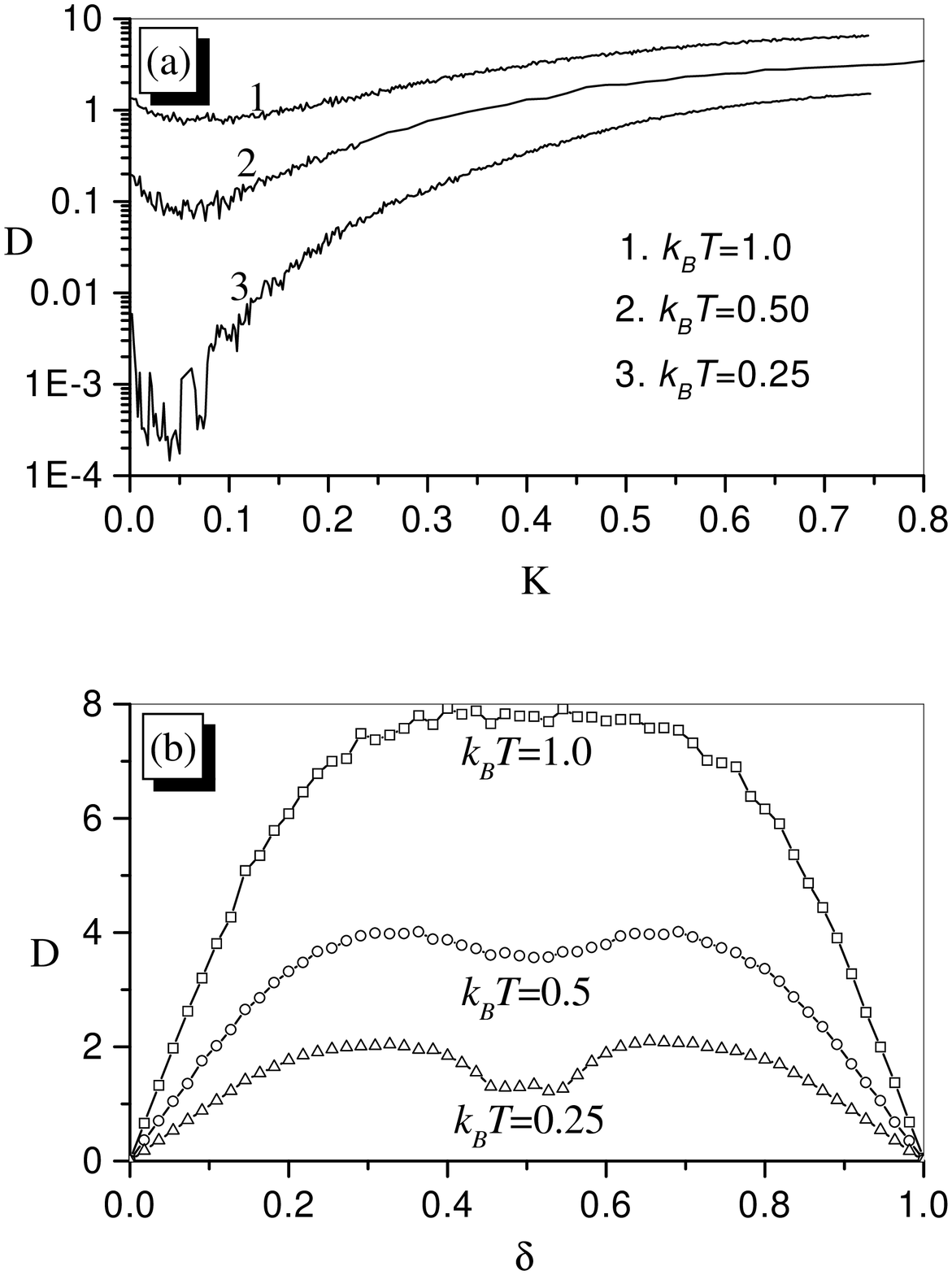}
\narrowtext
\caption{
(a). The diffusion coefficient $D$ of the FK chain versus the
coupling strength $K$ for different temperatures $k_BT=0.25$, $0.5$ and $1.0$. 
The vertical coordinate is shown by using a logarithm scale.
The diffusion coefficient first decreases against the coupling and then
increases and exceeds the value for the uncoupled case, indicating a
competition between order and disorder. For moderate coupling, the $ D-K$ relation
is a power type. (b). The diffusion rate $D$ against the winding number (frustration) 
$\delta $ of the chain for different temperatures $k_BT=0.25$, $0.5$ and $1.0$. 
The diffusion indicates a strong commensurability effect.
}
\end{figure}

\begin{figure}
\epsfxsize=8cm
\epsfbox{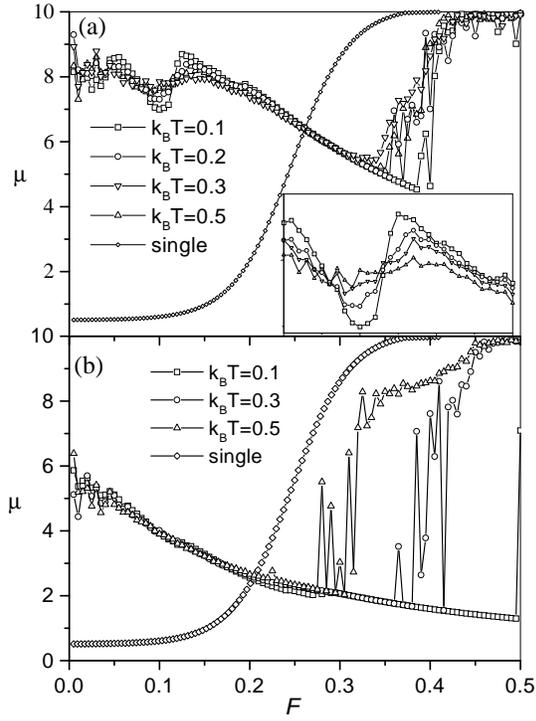}
\narrowtext
\caption{
The mobility of the chain $\mu $ varying with the external
driving force $F$ for cases (a): $N=8$, $M=3$ and (b): $N=8$, $M=1$ for
different temperatures. The comparison line corresponds to the case of a single
particle at the temperature $k_BT=0.5$. The inset of (a) is plotted to make a clearer
observation. In many regions the mobility decreases 
with increasing the external force for the coupled case. This anomalous behavior is the
consequence of the competition between the moving kink (traveling wave) and the phonons radiated
by the collision between the harmonic chain and the periodic potential.
}
\end{figure}

\begin{figure}
\epsfxsize=8cm
\epsfbox{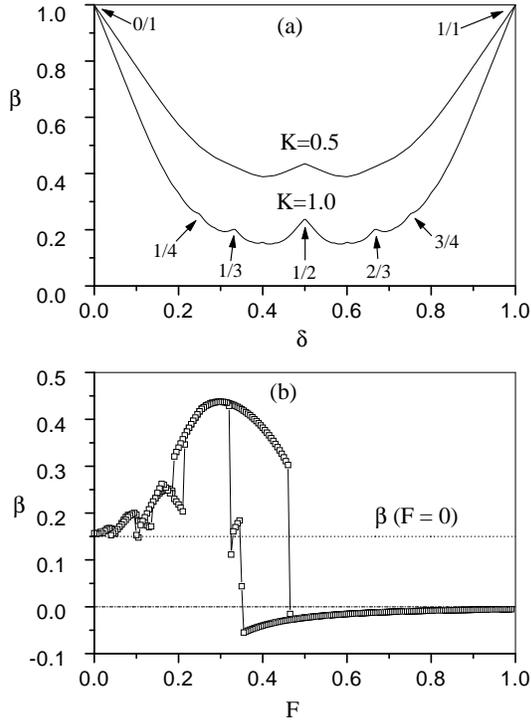}
\narrowtext
\caption{
(a): The contraction factor $\beta $ varies with the frustration 
$ \delta $ with $F=0$ for K=0.5 and 1.0. Peaks at the rational $\delta =0,1,1/2,...$ shows
the commensurate effect. (b): $\beta $ against the external force $F$ with 
$ \delta =3/8$. The dashed horizontal line corresponds to the value of $\beta $
at $F=0$. The hysteresis loop can be observed due to the bistability of the system.
}
\end{figure}

\begin{figure}
\epsfxsize=8cm
\epsfbox{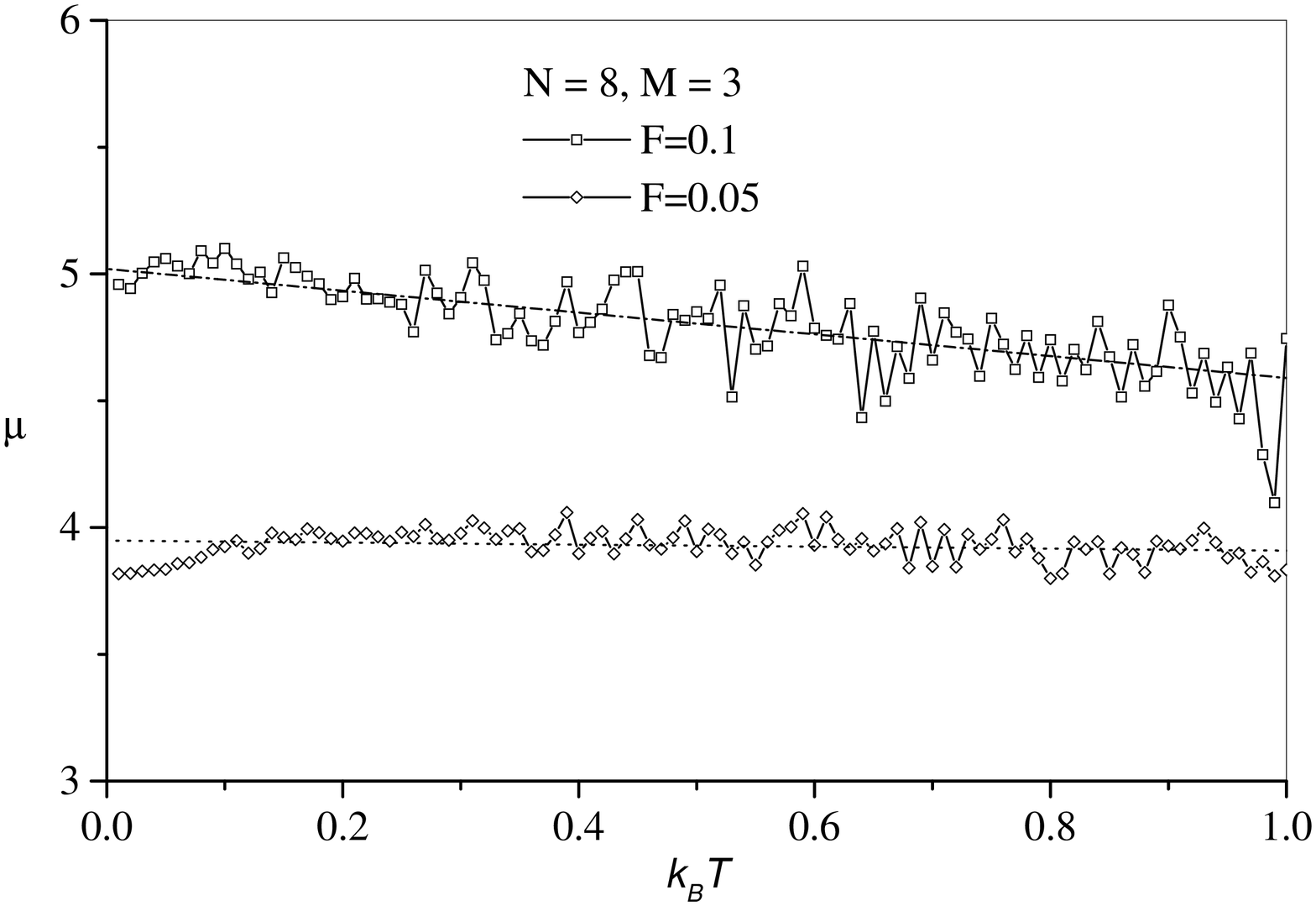}
\narrowtext
\caption{
The mobility of the chain $\mu $ varying with the the temperature $k_BT$ 
for the case $N=8,$ $M=3$ for different external drives $F=0.05$ and 
$0.1$. The mobility remains almost unchanged or decreases, which is much
different from the single-particle case.
}
\end{figure}


\begin{references}

\bibitem{risken} {H.Risken, The Fokker-Planck Equation (Springer-Verlag,
 New York, 1984).}

\bibitem{Ben} {E.Ben-Jacob, D.J.Bergman, B.J.Matkowsky, and Z.Schuss, Phys. Rev. {\bf A
26}, 2805 (1982).}

\bibitem{Schneider} {T.Schneider, E.P.Stoll, R.Morf, Phys. Rev. {\bf B18}, 1417 (1978);
T.Schneider and E.Stoll, Phys. Rev. Lett . {\bf 41}, 1429 (1978);
T.Schneider and E.Stoll, Phys. Rev. {\bf B 22}, 5317 (1980);T.Schneider and
E.Stoll, Phys. Rev. {\bf B 22}, 395 (1980); R.A.Guyer and M.D.Miller, Phys.
Rev. {\bf A17}, 1774 (1978).}

\bibitem{superion} {M.B.Salamon, Physics of Superionic Conductors (Springer-Verlag, New
York, 1979).}

\bibitem{halperin} {V.Ambegaokar and B.I.Halperin, Phys. Rev. Lett.
{\bf 22}, 1364 (1969).}

\bibitem{vortex} {Chen Baoxing and Dong Jinming, Phys. Rev. {\bf B 44}, 10206 (1991);
Zhigang Zheng and Gang Hu, Comm. Theor. Phys. {\bf 27}, 157 (1997); M.Buttiker, E.P.Harris, 
and R.Landauer, Phys. Rev. {\bf B 28}, 1268 (1983).}

\bibitem{strogatz} {S.H.Strogatz, Nonlinear dynamics and Chaos (Addison-Wesley, Reading, MV,
1994).}

\bibitem{2D} {B.N.J.Persson, Phys. Rev. {\bf B 48}, 18140 (1995); J. Chem. Phys. {\bf 103}, 
3849 (1995); Sliding Friction: Principles and Applications (Springer-Verlag, Heidelberg, 1998). }

\bibitem{zheng1} {Zhigang Zheng and Gang Hu, Phys. Rev. {\bf E52}, 109 (1995).}

\bibitem{FK} {J.Frenkel and T.Kontorova, Phys. Z. Sowjet. {\bf 13}, 1 (1938).}

\bibitem{selke} {W.Selke, Phys. Rep. {\bf 170}, 213 (1988).}

\bibitem{aubry} {S. Aubry, Phys. Rep. {\bf 103}, 12 (1984); M.Peyrard and S.Aubry, J.
Phys. {\bf C 16}, 1593 (1983).}

\bibitem{ac} {L.Floria and J.Mazo, Adv. Phys. {\bf 45}, 505 (1996).}

\bibitem{watanabe} {S.Watanabe, H.Zant, S.Strogatz and T.Orlando, Physica {\bf D97}, 429
(1996).}

\bibitem{marz} {M.Mazroui and Y.Boughaleb, Physica {\bf A 227}, 93 (1996).}

\bibitem{braun1} {O.M.Braun, T.Dauxios, M.V.Paliy and M.Peyrard, Phys. Rev. Lett. {\bf 78}
, 1295 (1997); O.M.Braun, A.R.Bishop and J.Roder, Phys. Rev. Lett. {\bf 79},
3692 (1997).}

\bibitem{braun2} {O.M.Braun, T.Dauxois, M.V.Paliy and M.Peyrard, Phys. Rev. {\bf B 54},
321 (1996).}

\bibitem{braiman} {Y.Braiman, F.Family and H.G.E.Hentschel, Phys. Rev. {\bf E 53}, R3005
(1996).}

\bibitem{ustinov} {A.Ustinov, M.Cirillo and B.Malomed, Phys. Rev. {\bf B 51}, 3081 (1995).}

\bibitem{zheng2} {Zhigang Zheng, Bambi Hu and Gang Hu, CNS preprint (1998); 
Comm. Theor. Phys. (1998) (submitted).}

\bibitem{zheng3} {Zhigang Zheng, Bambi Hu and Gang Hu, Phys. Rev. {\bf B 58}, 5453 
(1998).}

\bibitem{zheng4} {Zhigang Zheng, Bambi Hu and Gang Hu, Phys. Rev. {\bf E 57}, 1397 (1998).}

\end{references}
\end{document}